\documentclass[aps,preprint]{revtex4}%
\usepackage{amsfonts}
\usepackage{amsmath}
\usepackage{amssymb}
\usepackage{graphicx}%
\providecommand{\U}[1]{\protect\rule{.1in}{.1in}}

\begin{document}
\preprint{LPTh 10/05}
\title{Hydrogen atom in momentum space with a minimal length}
\author{Djamil Bouaziz}
\email{djamilbouaziz@mail.univ-jijel.dz}
\author{Nourredine Ferkous}
\affiliation{Laboratoire de Physique Th\'{e}orique (LPTh), Universit\'{e} de Jijel,
Bo\^{\i}te Postale 98, Ouled Aissa, 18000 Jijel, Algeria.}

\begin{abstract}
A momentum representation treatment of the hydrogen atom problem with a
generalized uncertainty relation,which leads to a minimal length $\left(
\Delta X_{i}\right)  _{\min}=\hbar\sqrt{3\beta+\beta^{\prime}}$, is presented.
We show that the distance squared operator can be factorized in the case
$\beta^{\prime}=2\beta$. We analytically solve the s-wave bound-state
equation. The leading correction to the energy spectrum caused by the minimal
length depends on $\sqrt{\beta}$. An upper bound for the minimal length is
found to be about $10^{-9}$ fm.

\end{abstract}
\endpage{ }
\maketitle

\section{\textbf{Introduction \ }}

Since Kempf and co-workers\textit{ }developed the theoretical framework of
quantum mechanics based on a generalized uncertainty relation, which implies
the existence of a minimal length, in a series of papers
\cite{Kempf1,Kempf11,Kempf2}, a lot of attention has been attracted to the
study of physical problems within this formalism, see, for instance, Refs.
\cite{chang,Benczik,brau2,Bouaziz,Harbach,Nouicer,bouaziz2,bina,quesne,merad}.
The idea of modifying the standard Heisenberg uncertainty relation in such a
way that it includes a minimal length has first been proposed in the context
of quantum gravity and string theory \cite{Garay,Amati}. It is assumed that
this elementary length should be on the scale of the Planck length of
$l_{p}=10^{-35}$m, below which the resolution of distances is impossible.

It was shown in Refs. \cite{Kempf1,Kempf11,Kempf2} that the minimal length
uncertainty relation is closely connected to a modification of the standard
Heisenberg algebra by adding specific corrections to the canonical commutation
relations between position and momentum operators, so that the Heisenberg
algebra becomes $[\widehat{X}_{i},\widehat{P}_{j}]=i\hbar\lbrack
(1+\beta\widehat{P}^{2})\delta_{ij}+\beta^{\prime}\widehat{P}_{i}\widehat
{P}_{j}]$, where $\beta$ and $\beta^{\prime}$ are small positive parameters
related to the minimal length by $\left(  \Delta X_{i}\right)  _{\min}%
=\hbar\sqrt{3\beta+\beta^{\prime}}$. One of the fundamental consequences of
the generalized uncertainty relation is the loss of localization in coordinate
space due to the presence of a nonzero minimal uncertainty in position
measurements. Consequently, momentum space is more convenient in order to
solve any eigenvalue problem. However, this is not often possible, especially
when the potential depends, in a not too straightforward manner, on the
position operators as in the case of the hydrogen atom potential. In the
literature, this problem is the most studied in this modified version of
quantum mechanics \cite{akhoury,brau,sandor,stetsko1,stetsko2}. This is
natural because this system has a particular interest. The elementary length
has been associated to with finite size of the electron; and the use of the
high-precision experimental data for the transition 1$S$ - 2$S$ and for the
Lamb shift were exploited to estimate an upper bound for the minimal length of
about $0.01-0.1$ fm.

Except in Ref. \cite{akhoury}, the energy spectrum of the hydrogen atom has
been obtained perturbatively in coordinate space: The terms proportional to
the deformation parameters $\beta$ and $\beta^{\prime}$ in the Schr\"{o}dinger
equation were regarded as perturbation corrections to the Hamiltonian
operator; the use of perturbation theory allowed for the computation of the
corrections in the first order to the energy levels. In momentum space, the
difficulty lies in defining the square root of the operator $\widehat{R}^{2}=%
{\textstyle\sum\limits_{i=1}^{3}}
\widehat{X}_{i}\widehat{X}_{i}$. To avoid this problem, the author of Ref.
\cite{akhoury} used complicated successive transformations on the wave
function and solved the $s$-wave bound-state equation. The spectrum is
obtained, however, the correction caused by the minimal length is different
from what was obtained in coordinate spaces.

Given the discrepancy between the results, which concern this problem, it is
interesting to consider it again with another method. For this purpose, here,
we give a simple method to solve the $s$-wave deformed Schr\"{o}dinger
equation in momentum space for the Coulomb potential. We show that, in the
particular case $\beta^{\prime}=2\beta$, the distance squared operator
$\widehat{R}^{2}$ can be factorized in the first order in the deformation
parameter $\beta$. We obtain the wave function and the energy spectrum, which
are different from that of Ref. \cite{akhoury}. By using the experimental data
for the Lamb shift, we find an upper bound of the minimal length of $10^{-9}$ fm.

The rest of this paper is organized as follows. In Sec. II, we give a brief
review of different works, which concern the hydrogen potential with a minimal
length. In Sec. III, we study this problem in momentum space. We summarize our
results in a brief concluding section.

\section{Hydrogen atom with a minimal length: A review}

As mentioned in Sec. I, several papers have been devoted to the study of the
hydrogen atom problem in quantum mechanics with a generalized uncertainty
relation, based on the following deformed Heisenberg algebra
\cite{akhoury,brau,sandor,stetsko1,stetsko2}:
\begin{align}
\left[  \widehat{X}_{i},\widehat{P}_{j}\right]   &  =i\hbar\left[  \left(
1+\beta\widehat{P}^{2}\right)  \delta_{ij}+\beta^{\prime}\widehat{P}%
_{i}\widehat{P}_{j}\right]  ,\\
\left[  \widehat{P}_{i},\widehat{P}_{j}\right]   &  =0,\\
\left[  \widehat{X}_{i},\widehat{X}_{j}\right]   &  =i\hbar\frac{2\beta
-\beta^{\prime}+\beta\left(  2\beta+\beta^{\prime}\right)  \widehat{P}^{2}%
}{1+\beta\widehat{P}^{2}}\left(  \widehat{P}_{i}\widehat{X}_{j}-\widehat
{X}_{i}\widehat{P}_{j}\right)  .
\end{align}
Many representations of the operators $\widehat{X}_{i}$ and $\widehat{P}_{i}$
were used by assuming that $\hat{X}_{i}$ and $\hat{P}_{i}$ are functions of
the operators $\hat{x}_{i}$ and $\hat{p}_{i}$, which satisfy the standard
canonical commutation relations of ordinary quantum mechanics.

Brau was the first to use the perturbation technique to calculate the
correction to the energy spectrum of the hydrogen atom due to the presence of
a minimal length\ \cite{brau}. The author made a simple choice of $\hat{X}%
_{i}$ and $\hat{P}_{i}$ in the coordinate space valid in the case
$\beta^{\prime}=2\beta$, and in the first order in $\beta$, namely
\begin{equation}
\widehat{X}_{i}=\widehat{x}_{i},\text{ \ \ }\widehat{P}_{i}=\widehat{p}%
_{i}\left(  1+\beta\widehat{p}^{2}\right)  . \label{brau}%
\end{equation}
Thus, the Schr\"{o}dinger equation takes the form%
\begin{equation}
\left(  \frac{\widehat{p}^{2}}{2m}+V(\widehat{\overset{\rightarrow}{r}}%
)+\frac{\beta}{m}\widehat{p}^{4}\right)  \psi(\overset{\rightarrow}{r}%
)=E\psi(\overset{\rightarrow}{r}).
\end{equation}

As is clearly seen, the effect of the minimal length is described by the
presence of a perturbation term ($\frac{\beta}{m}\widehat{p}^{4}$) in the
ordinary Schr\"{o}dinger equation.

Thereafter, Akhoury and Yao \cite{akhoury} considered the same problem in
momentum space by using the following representation:%
\begin{equation}
\widehat{X}_{i}=\left(  1+\beta\widehat{p}^{2}\right)  \widehat{x}_{i}%
+\beta^{\prime}\widehat{p}_{i}\widehat{p}_{j}\widehat{x}_{j}+\gamma\widehat
{p}_{i},\text{ \ \ \ }\widehat{P}_{i}=\widehat{p}_{i}. \label{Akhory}%
\end{equation}

The authors write the Schr\"{o}dinger equation for the Coulomb potential
$V(r)=-\alpha/r$, in the form%
\[
\left(  \widehat{R}\left(  \frac{\widehat{p}^{2}}{2m}-E\right)  -\alpha
\right)  \left\vert \psi\right\rangle =0,
\]
where $\widehat{R}$ is the square root of the operator $\widehat{R}^{2}=%
{\textstyle\sum\limits_{i=1}^{3}}
\widehat{X}_{i}\widehat{X}_{i}$.

Unlike in ordinary quantum mechanics, where the expression of $\widehat{R}$
can be obtained for the $s$-waves ($l=0$), the definition of this operator is
not obvious in the deformed case even if $l=0$.

To overcome this problem, Akhoury and Yao performed some changes of variables
and transformations on the wave function, and defined a supposedly radial
distance operator $\widehat{R}=i\hbar\lbrack1+\left(  \beta+\beta^{\prime
}\right)  p^{2}]\dfrac{d}{dp}$, which acts on the state $\tau\left\vert
\psi\right\rangle $ instead of $\left\vert \psi\right\rangle $, where $\tau$
is a transformation not explicitly given. Nevertheless, the authors succeeded
to get a solution to the deformed Schr\"{o}dinger equation, and to extract the
energy spectrum by imposing the condition of single valuedness on the wave
function. The correction to the energy levels is completely
different\textit{\ }from that obtained by Brau. It is important to mention
that in Ref. \cite{akhoury}, the condition of single valuedness was not
correctly applied. In Sec. III, we propose another method in momentum space;
the correct energy spectrum will be calculated.

The problem of the hydrogen atom has been reconsidered by Benczik \textit{et}
\textit{al.} \cite{sandor}, by using the representation given by Eq.
(\ref{Akhory}) with two approaches, the first by numerical techniques in
momentum space and the second by the perturbation theory in position space.
Their results are in disagreement with those obtained by Akhoury and Yao, and
differ from the ones of Brau only for $\ell=0$.

Finally, Stetsko and Tkachuk \cite{stetsko1} and \cite{stetsko2} proposed
another perturbative method by using the following position representation of
the operators $\widehat{X}_{i}$ and $\widehat{P}_{i}$ :
\begin{equation}
\widehat{X}_{i}=\widehat{x}_{i}+\frac{2\beta-\beta^{\prime}}{4}\left(
\widehat{p}^{2}\widehat{x}_{i}+\widehat{x}_{i}\widehat{p}^{2}\right)  ,\text{
\ \ \ \ }\widehat{P}_{i}=\widehat{p}_{i}\left(  1+\frac{\beta^{\prime}}%
{2}\widehat{p}^{2}\right)  . \label{Stetsko}%
\end{equation}
This representation reduces to that of Brau Eq. (\ref{brau}) in the case
$\beta^{\prime}=2\beta$. They compute the correction to the energy spectrum in
the first order in $\beta$ and $\beta^{\prime}$. Their results reproduce those
of \ Brau even in the case $\ell=0$.

For the sake of completeness, let us mention that the Coulomb potential has
also been considered in the one-dimensional case in both nonrelativistic
\cite{fityo} and relativistic \cite{chargui} quantum mechanics with a minimal
length. The treatment was performed in momentum space, and the expression of
the energy spectrum is different from that obtained in the case $\ell=0$ of
Ref. \cite{akhoury}.

In the following, we consider, again, the hydrogen atom problem by using a
momentum representation, which is more appropriate in this version of quantum mechanics.

\section{Momentum space treatment}

Let us consider the Schr\"{o}dinger equation for the hydrogen atom in the form%
\begin{equation}
\left(  \widehat{R}(\frac{\widehat{p}^{2}}{2m}-E)-\alpha\right)  \left\vert
\psi\right\rangle =0, \label{Sh}%
\end{equation}
where the strength of the potential is $\alpha=\frac{e^{2}}{4\pi\epsilon_{0}}%
$. In the momentum representation, the wave function reads \cite{chang}%
\[
\psi(\overset{\rightarrow}{p})=\langle\overset{\rightarrow}{p}\left\vert
\psi\right.  \rangle=Y_{\ell}^{m}(\theta,\varphi)\psi(p).
\]

By restricting ourselves to the $\ell=0$ wave function and by using the
momentum representation given by Eq. (\ref{Akhory}), with $\gamma=0$, we
obtain the following expression for the distance squared operator :%
\begin{equation}
\widehat{R}^{2}\mathbf{=}\left(  i\hbar\right)  ^{2}\left\{  \left[  1+\left(
\beta+\beta^{\prime}\right)  p^{2}\right]  ^{2}\frac{d^{2}}{dp^{2}}+\frac
{2}{p}\left[  1+\left(  \beta+\beta^{\prime}\right)  p^{2}\right]  \left[
1+\left(  2\beta+\beta^{\prime}\right)  p^{2}\right]  \frac{d}{dp}\right\}  .
\label{R2}%
\end{equation}

In the general case, this operator is not factorizable in the sense that its
square root is unknown. In spite of this, we show that $\widehat{R}^{2}$ can
be factorized in the particular case $\beta^{\prime}=2\beta$ in the first
order\ in $\beta$. Indeed, since $\beta$ and $\beta^{\prime}$ are supposed to
be small parameters, the distance squared operator can be expressed as%
\begin{equation}
\widehat{R}^{2}\mathbf{=}\left(  i\hbar\right)  ^{2}\left\{  (1+6\beta
p^{2})\frac{d^{2}}{dp^{2}}+\frac{2}{p}\allowbreak(1+7\beta p^{2})\frac{d}%
{dp}\right\}  +O\left(  \beta^{2}\right)  . \label{R1}%
\end{equation}
In Eq. (\ref{R1}), $\widehat{R}^{2}$ can be written as $\widehat{R}%
\times\widehat{R}$\textbf{,} where%
\begin{equation}
\widehat{R}\mathbf{=}i\hbar\left[  \left(  1+3\beta p^{2}\right)  \frac{d}%
{dp}+\frac{1}{p}\allowbreak\allowbreak\left(  1+\beta p^{2}\right)  \right]
+O\left(  \beta^{2}\right)  . \label{R}%
\end{equation}
From Eqs. (\ref{Sh}) and (\ref{R}), the radial Schr\"{o}dinger equation for
the hydrogen atom in momentum space with a minimal length reads%
\begin{equation}
\left(  1+3\beta p^{2}\right)  \left(  p^{2}+k^{2}\right)  \frac{d\psi(p)}%
{dp}+\left\{  \frac{1}{p}\allowbreak\allowbreak\left(  1+\beta p^{2}\right)
\left(  p^{2}+k^{2}\right)  +2p\left(  1+3\beta p^{2}\right)  +\frac{2i\alpha
m}{\hbar}\right\}  \psi(p)=0,
\end{equation}
where $k^{2}=-2mE$.

In order to integrate this equation, it is convenient to make it in the form%
\begin{equation}
\frac{d\psi(p)}{dp}+\left\{  \frac{\eta-1}{p+ik}\allowbreak\allowbreak
-\frac{\eta+1}{p-ik}+\frac{\frac{1}{3}-\xi}{p+i/\sqrt{3\beta}}+\frac{\frac
{1}{3}+\xi}{p-i/\sqrt{3\beta}}-\frac{1}{p}\right\}  \psi(p)=0, \label{dif}%
\end{equation}
in which,%
\[
\xi=\frac{\alpha m\sqrt{3\beta}}{\hbar\left(  1-3\beta k^{2}\right)  },\text{
\ \ \ }\eta=\frac{\alpha m}{\hbar k\left(  1-3\beta k^{2}\right)  }.
\]

The solution to Eq. (\ref{dif}) is%

\begin{equation}
\psi(p)=A\frac{\left(  1+3\beta p^{2}\right)  ^{1/3}}{p\left(  p^{2}%
+k^{2}\right)  }\exp\left[  2\xi i\arctan\left(  p\sqrt{3\beta}\right)  -2\eta
i\arctan\left(  p/k\right)  \right]  , \label{wf}%
\end{equation}
where $A$ is a normalization constant. In the limit $\beta=0$, $\psi(p)$
reduces to the result of ordinary quantum mechanics \cite{eugene}.

Our wave function differs from that obtained in Ref. \cite{akhoury} by the
factor $\frac{1}{p}\left(  1+3\beta p^{2}\right)  ^{1/3}$:%
\[
\psi(p)=\frac{1}{p}\left(  1+3\beta p^{2}\right)  ^{1/3}\psi
(p)_{\text{Akhoury}}.
\]
This discrepancy is due to a certain transformation used in Ref.
\cite{akhoury}, which has not been explicitly given.

To extract the energy spectrum, by following Refs.
\cite{akhoury,lombardi,eugene,n,nieto}, we require that $\psi(p)$ must be a
single-valued function ( i.e., it must be unchanged under the transformation):%
\[
\arctan\left(  z\right)  \rightarrow\arctan\left(  z\right)  +\pi.
\]
Thus, we must have%
\begin{equation}
\xi-\eta=n, \label{sp}%
\end{equation}
where $n$ is an integer number. This leads to the following quantization
condition :%
\begin{equation}
\frac{\alpha m}{\hbar k\left(  1+k\sqrt{3\beta}\right)  }=n,\text{
\ \ }n=1,2,...\text{ .} \label{Quant}%
\end{equation}

By solving for\ $k$ and by using $k=\sqrt{-2mE}$, we obtain%
\[
E_{n}^{\pm}=-\frac{1}{24m\beta}\left(  1\pm\left(  1+4\frac{m\alpha}{\hbar
n}\sqrt{3\beta}\right)  ^{1/2}\right)  ^{2}.
\]

In the limit $\beta\rightarrow0$, $E_{n}^{+}$ diverges. So, the energy
spectrum reads
\begin{equation}
E_{n}=-\frac{1}{24m\beta}\left(  1-\left(  1+4\frac{m\alpha}{\hbar n}%
\sqrt{3\beta}\right)  ^{1/2}\right)  ^{2},\ \text{\ \ \ \ }n=1,2,... \label{s}%
\end{equation}

To leading orders in the small parameter $\beta$, the spectrum can be
expressed as follows:%
\begin{equation}
E_{n}=-\left(  \frac{m\alpha^{2}}{2\hbar^{2}n^{2}}-\frac{m^{2}\alpha^{3}%
}{\hbar^{3}n^{3}}\sqrt{3\beta}+\frac{15}{2}\frac{m^{3}\alpha^{4}}{\hbar
^{4}n^{4}}\beta\right)  +O\left(  \beta^{\frac{3}{2}}\right)  \allowbreak
,\ \text{\ \ \ }n=1,2,... \label{spectrum}%
\end{equation}

The first term represents the energy spectrum of ordinary quantum mechanics,
while the second and third terms are the corrections brought about by the
existence of a minimal length. As we see, our result coincides with that of
the one-dimensional Coulomb potential \cite{fityo}, where the quantization
condition has been derived by imposing the Hermiticity of the Hamiltonian.

The main feature of the spectrum Eq. (\ref{spectrum}) is the presence of a
positive correction proportional to the minimal length ($\left(  \Delta
X\right)  _{\min}=\hbar\sqrt{3\beta}$), which is the leading correction.
Previously, this term was omitted in Ref. \cite{akhoury}; the correction due
to the modification of the Heisenberg algebra is negative, and is described
only by the third term of Eq. (\ref{spectrum}). This is because the condition
of the single valuedness was not strictly applied. The authors took, instead
of Eq. (\ref{sp}), the condition $\eta=n$, which is not sufficient to assure
that the wave function be single valued. In the perturbative treatment of the
hydrogen atom \cite{brau,sandor,stetsko1,stetsko2}, $\beta$ is the
perturbation parameter, and, naturally the first-order correction is
proportional to this deformation parameter.

This result is very important because it leads to an order of magnitude of the
minimal length completely different from what was obtained in Refs.
\cite{akhoury,brau,sandor,stetsko1,stetsko2}, where an upper bound for the
minimal length was found to be about $0.01-0.1$ fm. The estimation of this
bound was mainly obtained by two methods. The first requires that the
corrections to the spectrum due to the modification of the Heisenberg algebra
are smaller than the experimental error on the value of the transition $1S-2S$
in the hydrogen atom \cite{brau,akhoury}. The second assumes that the effects
of the minimal length are included in the gap between the theoretical and the
experimental values of the Lamb shift for the hydrogen atom levels
\cite{sandor,stetsko1,stetsko2}.

Indeed, we now use Eq. (\ref{spectrum}) to give a new constraint for the
minimal length. Let us write, to leading orders in the small parameter $\beta
$, the following relative shift:
\begin{equation}
\frac{E_{2S}-E_{1S}}{E_{1S}}=\allowbreak-\frac{3}{4}+\frac{m\alpha}{4\hbar
}\sqrt{3\beta}-\frac{21}{16}\frac{m^{2}\alpha^{2}}{\hbar^{2}}\beta+O\left(
\beta^{\frac{3}{2}}\right)  . \label{cc}%
\end{equation}

The $1S-2S$ energy splitting in the hydrogen atom is measured with an accuracy
of $\varepsilon=1.8\times10^{-14}$ \cite{Niering}. If we attribute this error
entirely to the minimal length correction (\ref{cc}), and by taking only the
first dominant contribution intoaccount, we can write%

\[
\varepsilon=\frac{m\alpha}{4\hbar}\sqrt{3\beta}+O\left(  \beta\right)  ,
\]
which gives the value $\left(  \Delta X\right)  _{\min}\sim5\times10^{-9}$ fm.
This upper bound is much smaller than the one obtained in Refs.
\cite{akhoury,brau}. This is due to the absence of the term proportional to
$\sqrt{3\beta}$ in these references.

It was noted in Ref. \cite{sandor} that a better estimate for the minimal
length is obtained by including its corrections in the Lamb shift. Thus, from
Eq. (\ref{spectrum}), the difference $\left(  L_{1S}^{\exp}-L_{1S}%
^{th}\right)  $ can be taken as%

\[
\Delta E_{1S}=\frac{m^{2}\alpha^{3}}{\hbar^{3}n^{3}}\sqrt{3\beta}=h\left(
L_{1S}^{\exp}-L_{1S}^{th}\right)  .
\]

Given $L_{1S}^{th}=8172,731(40)$ MHz \cite{Mallampalli} and $L_{1S}^{\exp
}=8172,837(22)$ MHz \cite{Schwob}, we obtain the value $\left(  \Delta
X\right)  _{\min}\sim10^{-6}$ fm. Again, we have a stringent limit on the
value of the minimal length. One can conclude that, in non relativistic
treatment of the hydrogen atom problem, the inclusion of a minimal length
would not affect the hydrogen atom physics because its predicted size is too small.

\section{\bigskip Summary}

We have proposed a simple method to solve the $s$-wave Schr\"{o}dinger
equation in momentum space for the hydrogen atom problem in the framework of
quantum mechanics with a generalized uncertainty relation, characterized by a
minimal length $\left(  \Delta X\right)  _{\min}=\hbar\sqrt{3\beta
+\beta^{\prime}}$. We have shown that the distance squared operator
$\widehat{R}^{2}$\ is factorizable in the case $\beta^{\prime}=2\beta$ in the
first order\ in the deformation parameter $\beta$. The wave functions and the
corresponding energy levels are obtained. The leading correction to the energy
spectrum is proportional to $\sqrt{\beta}$, which is in agreement with that of
the one-dimensional case \cite{fityo}. The dependence on $\sqrt{\beta}$
drastically lowers the minimal length scale, which is of about $10^{-9}$ fm.
This leads us to conclude that the minimal length in the problem considered
here is too small so that its effects on the hydrogen atom physics are negligible.

\begin{acknowledgments}
We thank Professor Tahar Boudjedaa for several very instructive discussions.
This research is supported by the Algerian Ministry of Higher Education and
Scientific Research under the CNEPRU Projects No D01720060022 and No D01720090023.
\end{acknowledgments}

\end{document}